# A comparison of proton ranges in complex media using GATE/Geant4, MCNP6 and FLUKA

Jarle Rambo Sølie[a,b,*], Helge Egil Seime Pettersen[b,c], Ilker Meric[a], Odd Harald Odland[c], Håvard Helstrup[d], Dieter Röhrich[b]

* Corresponding author: E-mail: jars@hvl.no

[a]Department of Electrical Engineering, Western Norway University of Applied Sciences, Inndalsveien 28, 5063 Bergen, Norway
[b]Department of Physics and Technology, University of Bergen, Allégaten 55, 5007 Bergen Norway
[c]Department of Oncology and Medical Physics, Haukeland University Hospital, Jonas Lies vei 65, 5021 Bergen, Norway
[d]Department of Computing, Mathematics and Physics, Western Norway University of Applied Sciences, Inndalsveien 28, 5063 Bergen, Norway

## Abstract

The Monte Carlo simulation method is a powerful tool for radiation physicists, and several general-purpose software packages are commonly applied in a myriad of different radiation physics fields today. In medical physics, charged particle detectors for proton Computed Tomography are under development, a modality introduced in order to increase the accuracy of proton radiation therapy. Monte Carlo simulations are helpful during the development and optimization phase of such detector systems, when used for construction of analysis software in order to predict system performance, and for the optimization phase by way of comparing the predicted performance of proposed prototype systems. In order to justify the usage of Monte Carlo for such purposes, the simulation output must be validated against experimental or theoretical data, or even cross-checked between different Monte Carlo software packages. A comparative Monte Carlo validation will increase confidence in the applied Monte Carlo software packages and its package-specific algorithms, user-customizable settings and implemented physics models. In this study, we compare three general-purpose Monte Carlo software packages (GATE/Geant4, MCNP6 and FLUKA) with respect to how they predict the spatial distribution of the stopping position of protons. They are compared to each other and to semi-empirical data using the mean proton range, the longitudinal and lateral variation of individual proton ranges, respectively called beam straggling and beam spreading, and the fraction of primary protons lost to nuclear interactions. This comparison is performed in two homogeneous materials and in a detector geometry designed for proton Computed Tomography. The three Monte Carlo software packages agree well, and sufficiently reproduce the semi-empirical data. Some discrepancies are observed, such as less lateral beam spreading in GATE/Geant4, and a small deficiency in the MCNP6 proton range in water: This is consistent with previously published data. Due to the general agreement, the choice of simulation framework may be made on personal preferences or inter-project compatibility. It is important to note that the choice of physics packages, simulation parameter settings and material definitions are important aspects when performing Monte Carlo simulations, both during the preparation, execution and interpretation of the simulation results.

# 1 Introduction

The Monte Carlo (MC) simulation method is at present a common, powerful and versatile tool widely used in physics research where the study of interactions between ionising radiation and matter is of importance. MC simulation can be a valuable tool during the development and design phases of detectors due to its ability to assess the optimal design parameters prior to experimental efforts. A good example of such applications of MC simulations is within proton Computed Tomography (proton CT). The detector systems required for proton CT are technically challenging, complex and costly to build. Thus, MC simulations have been extensively used for the design and optimization of such detectors (Giacometti et al., 2017; Lee et al., 2016; Steinberg et al., 2012).

The Bergen proton CT group is currently conducting research with the aim of developing a digital tracking calorimeter for the purposes of performing proton CT scans (Pettersen et al., 2017). The goal is to develop a detector system with the ability to track individual protons and find the residual ranges so that a three-dimensional proton stopping power map of the traversed matter can be reconstructed which, in turn, can be used in dose planning for proton radiation therapy, resulting in a higher treatment accuracy compared to standard methods. Thus, the technical requirements are stringent and a systematic sub-millimetre precision in the determination of proton ranges is required (Poludniowski et al., 2015; Sadrozinski et al., 2004).

Preliminary design work such as deciding upon the type of detector, number of detector layers and type of absorber material is currently underway. Further detector optimization will also be carried out with the aid of MC simulations. In addition, it should also be noted that the required range-energy tables for the detector geometry will be generated using MC simulations. The fact that range-energy tables are generated through MC simulations in conjunction with the need for a systematic sub-millimetre precision makes it important to assess the potential differences between different MC software packages used for simulations of proton CT scans.

The study at hand aims therefore at applying different MC packages for calculating parameters relevant to a proton CT system such as proton range resolution and proton track reconstruction efficiency (Pettersen et al., 2017), and then comparing the results from the different MC packages. The relevant parameters are the longitudinal and lateral variation of individual proton ranges, respectively called beam straggling and beam spreading (from multiple Coulomb scattering), and the fraction of primary protons lost to nuclear interactions. Three general-purpose MC software packages are using to this end: "Geant4 Application for Emission Tomography" (GATE) (Agostinelli et al., 2003; Jan et al., 2004) "Monte Carlo N-Particle" (MCNP6) (Goorley et al., 2013) and "FLuktUierende KAskade" (FLUKA) (Ferrari et al., 2005).

Mono-energetic proton beams with energies in the therapeutic span of 50 – 230 MeV are simulated as they propagate through homogeneous water and aluminium phantoms in addition to a relatively complex proton tracking detector geometry containing an array of different materials. Semi-empirical data of proton ranges in water and aluminium from PSTAR (Berger et al., 2005), as well as data on the beam spreading and the amount of nuclear interactions from J.F. Janni (Janni, 1982), are included in the comparisons where applicable. The analysis of the MC simulated data is carried out using the ROOT data analysis framework (Brun and Rademakers, 1997). The analysis code used in this work is made freely available as a GitHub repository (Pettersen, 2017).

There exist numerous studies with focus on detailed comparisons between the pertinent MC packages. Kimstrand et al. (Kimstrand et al., 2008) have modelled and compared the transport of protons grazing a tungsten block by using Geant4.8.2, FLUKA2006 and MCNPX2.4.0 and found that, while the energy spectrum of out-scattered protons agreed between the MC software packages, dose-weighted out-scatter probability was highly dependent on user-defined settings, and quantitatively the deviation in out-scatter probability between simulations could reach up to 37%.

Other studies have shown discrepancies in beam spreading between different MC packages and experimental data, with Grevillot et al. (Grevillot et al., 2010) reporting that GATE/Geant4 underestimates the transversal spread, attributed to the multiple scattering (MS) model applied in GATE/Geant4. Bednarz et al. (Bednarz et al., 2011) report discrepancies in the multiple coulomb scattering algorithms between MCNPX and Geant4, with Geant4 being more accurate in calculating scattering angle and MCNPX being more accurate in calculating displacement when compared to theories of Moliere and Highland. Mertens et al. (Mertens et al., 2010) notes that MCNP overestimates the spread in low density and low-Z targets, suggesting inaccuracies in the scattering cross-sections as a reason for the overestimation. Recently, Lin et al (Lin et al., 2017) have investigated the angular distributions of protons after hitting water and aluminium targets, as well as the Bragg peak position in a water phantom, and found similar inconsistencies in the lateral beam spread, but good agreement of longitudinal Bragg peak positions.

It has come to the attention of the authors that, while the literature concerning comparisons of MC packages is extensive, there is little published information available on direct comparisons of proton ranges and range straggling between MC packages at therapeutic energies. Also, to the best of the authors' knowledge, the results of the present work represent a first attempt to compare proton ranges and range straggling calculated by the above-mentioned MC packages in a heterogeneous, layered calorimeter geometry.

In the remainder of this work, a brief description of the physics settings applied in the individual MC packages and a presentation of the materials and geometries applied in the MC simulations will be given. This will be followed by a presentation and comparison of the results obtained by the three MC packages. Finally, the results will be discussed and followed by our conclusions from this work.

# 2 Material and methods

Three MC software packages GATE 7.2/Geant4 10.2.2, MCNP6.1 and FLUKA 2011.2c.5 were used to simulate monoenergetic proton beams with energies between 50 – 230 MeV, in 10 MeV increments, as they propagate and come to a complete stop inside different geometry such as a homogeneous water phantom, homogeneous aluminium phantom and the modelled proton tracking detector geometry. The pertinent detector geometry is shown in **Figure 1**.

In the simulations of the above-mentioned geometries, the incident proton beam was defined as a point source-beam starting 1 mm proximal to the front surface of the phantoms, and consisted of $10^5$ primary protons for each of the energies used. The physics packages chosen for each MC package, ensuring that the relevant physics processes and thresholds are accounted for in the simulations, are listed in **Table 1**.

For GATE/Geant4, the physics builder list QGSP_BIC_EMY is applied as recommended for MC simulations in proton therapy and proton imaging due to a variable maximum allowed simulation step size decreasing towards the Bragg Peak, and a high resolution binning of the pre-calculated stopping power tables (Grevillot et al., 2010; Z. Jarlskog and Paganetti, 2008). In MCNP6, nuclear interactions were modelled using the Cascade Exciton Model (CEM) 03.03 which is the recommended model for nuclear interactions (Goorley et al., 2013). Use of tabulated cross-sectional data was turned off and nuclear interactions were treated using only interaction models. The default Vavilov model for charged particle straggling was used and for multiple scattering, the default FermiLab angular deflection model with Vavilov straggling (Mokhov and Striganov, 2002) was used. It should be noted that all simulations using MCNP6 were run in the "proton-only" mode, thus ignoring the transport of all secondary particles other than protons. For FLUKA the predefined physics setting "PRECISIO" is recommended for precision simulations with respect to transport thresholds and activation of processes as detailed in the FLUKA manual (Ferrari et al., 2005). It is important to note that a manual adjustment of the ionization potentials for the different materials is possible in both GATE/Geant4 and FLUKA, whereas changing the automatically set, or calculated, ionization potentials in MCNP6 requires access to the source code (James, 2016; Seravalli et al., 2012).

The homogeneous water and aluminium phantoms were defined to have a cross sectional area of 10 x 10 $cm^2$ and was 40 cm in length, thus stopping all primary protons with energies up to 230 MeV. The geometry and material definition of the individual layers in the detector geometry is described in **Table 2**.

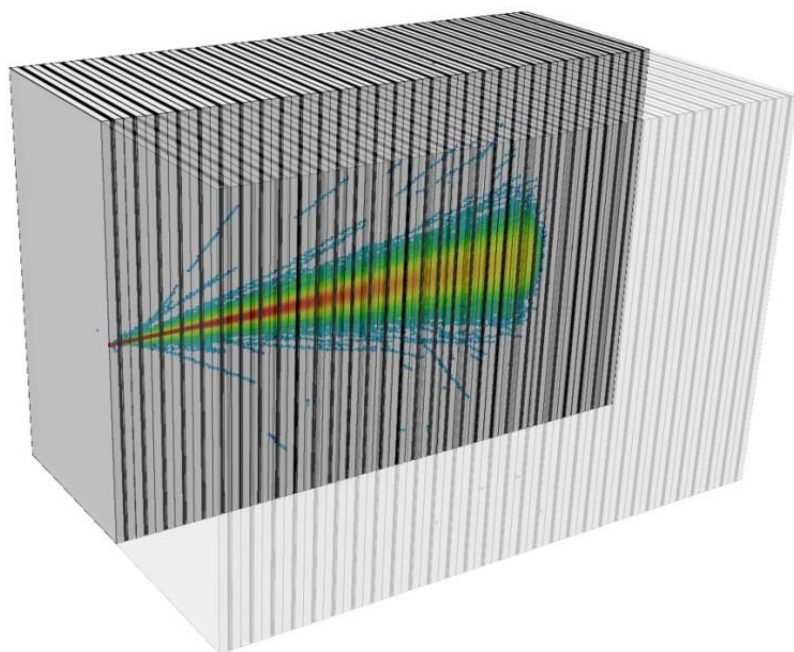

*Figure 1* The proton tracking detector geometry consisting of 30 layers is overlaid with a MC simulated primary proton beam consisting of $10^5$ protons inside the middle of the detector.

*Table 1.* The applied physics packages and parameters of the MC software packages considered in this work.

| MC package | Applied physics package | Parameters/notes |
|---|---|---|
| **GATE/ Geant4** | QGSP_BIC_EMY: Using the "option 3" electromagnetic model (Grevillot et al., 2010; Z. Jarlskog and Paganetti, 2008) | Mean Ionization potential for water manually set to 75 eV to match PSTAR data tables (Berger et al., 2005). |
| **MCNP6** | Cascade Exciton Model (CEM) for nuclear interactions. Vavilov straggling model for charged particle straggling (Mokhov and Striganov, 2002). | Mean Ionization potential for water is automatically set to 75 eV by MCNP6, otherwise, Bragg additivity is used to calculate its value for mixtures and compounds (James, 2016; Seravalli et al., 2012). |
| **FLUKA** | PRECISIO (Ferrari et al., 2005) | Particle transport threshold set at 100 keV. Mean Ionization potential for water manually set to 75 eV. |

*Table 2 Description of the geometry representing the proton tracking detector geometry. A single layer is modelled as 10 x 10 cm^2 and 4.975 mm thick slab made up of each of the materials listed below in the given order. This is repeated 30 times to obtain the complete detector geometry.*

| Slab element name | Material | Thickness [µm] |
|---|---|---|
| Absorber | Aluminium | 2000 |
| PCB Glue | Silver glue | 40 |
| PCB | Quartz epoxy | 160 |
| Chip glue | Silver glue | 40 |
| Passive chip | Silicon | 106 |
| Active chip | Silicon | 14 |
| Air gap | Air | 170 |
| Filler absorber | Aluminium | 300 |
| Filler glue | Cyanoacrylate | 70 |
| Absorber | Aluminium | 2000 |
| Air gap between layers | Air | 75 |

This detector geometry, with a final cross-sectional area of 10x10 cm$^2$ and 15 cm in length, ensures that approximately all protons with energies up to 210 MeV will stop inside the detector, this is per the completed preliminary detector geometry design work (Pettersen et al., 2017). Note that while the highest energies are unavailable in this geometry configuration, this limitation in range should not qualitatively affect the outcome of this study.

The final coordinates of all primary protons that stop inside the phantoms and in the detector geometry were stored, essentially giving the range of each individual proton in their respective geometries. This distribution of ranges was subsequently analysed in ROOT through a Gaussian fitting procedure to obtain the mean range of the proton beam (Pettersen, 2017). The mean ranges in water and aluminium were then compared to the projected ranges from the "Stopping-power and range tables for protons" (PSTAR) range-energy database (Berger et al., 2005), and the respective range deviations were calculated. For the detector geometry, the range deviation is calculated as the difference between the range and the average results from the three MC packages as no accurate experimental values are available.

The range straggling, defined as the standard deviation of the range distribution, is obtained from and compared between the three MC packages. In the case of water and aluminium phantoms the MC calculated data are compared to that of J.F. Janni (Janni, 1982).

As detailed in the work done by Makarova et al. (Makarova et al., 2016), the so-called transverse beam spread can be calculated as the root mean square (RMS) value of the lateral distribution of the proton Bragg Peak positions ($\sigma_x$), divided by the corresponding proton range.

The fraction of nuclear interactions was obtained by collating the interaction-type metadata available in the different MC output files, which was then compared with the semi-empirical data from J.F. Janni (Janni, 1982) where applicable. A rule-of-thumb is that approximately 1% of the protons undergo nuclear interactions per cm of water, or 1% per cm Water Equivalent Thickness (WET) in materials other than water (Durante and Paganetti, 2016).

# 3 Results

Values for proton ranges, range straggling, beam spread and fraction of nuclear interactions were obtained through simulations with the three MC packages for the three different geometries: water, aluminium and the detector geometry.

## 3.1 Proton ranges

**Table 3** lists the MC simulated proton ranges of a few selected initial primary proton energies as well as corresponding data from PSTAR (Berger et al., 2005).

As is shown in **Figure 2 (a) – (c)**, the largest mean projected range deviation between the MC packages is less than 1.7 mm (0.5%) of the range as listed in PSTAR (Berger et al., 2005) for water and 0.2 mm (0.13%) for aluminium. Range deviation in the detector geometry, calculated as the deviation from the average of the ranges from the three MC packages, deviates no more than 0.2 mm (0.15%). It is noted that while FLUKA and GATE/Geant4 match each other well in water, MCNP6 yield a larger range deviation with increasing initial proton energy.

## 3.2 Proton range straggling

The obtained results for range straggling for some selected primary proton energies are listed in **Table 4** and complete MC simulation results are displayed in **Figure 2 (d) – (f)**.

All three MC packages show a similar amount of range straggling, with a maximum difference between the MC packages of 0.48 mm (12.5%) in water and 0.08 mm (4.5%) in aluminium. The tendency of increasing deviation with higher energies are observable in the results for the detector geometry, where the largest deviation is less than 0.24 mm (13.7%). Note that a higher variation in the range straggling depending on the initial proton energy is observed in the detector geometry, and this may explain the large variation between the MC packages.

*Table 3 Simulated MC ranges and PSTAR data for 50, 100, 150 and 230 MeV primary proton energies in the water phantom, aluminium phantom and detector geometry.*

| Material | Energy [MeV] | GATE/Geant4 [mm] | MCNP6 [mm] | FLUKA [mm] | PSTAR [mm] |
|---|---|---|---|---|---|
| Water | 50 | 22.2 | 22.2 | 22.2 | 22.2 |
| | 100 | 77.0 | 76.8 | 77.0 | 77.1 |
| | 150 | 157.3 | 156.9 | 157.3 | 157.6 |
| | 230 | 328.7 | 327.4 | 328.6 | 329.1 |
| Aluminium | 50 | 10.8 | 10.8 | 10.9 | 10.8 |
| | 100 | 37.0 | 36.9 | 37.1 | 37.0 |
| | 150 | 75.0 | 75.0 | 75.3 | 75.1 |
| | 230 | 155.8 | 156.1 | 156.3 | 156.0 |
| Detector geometry | 50 | 11.1 | 11.1 | 11.1 | - |
| | 100 | 37.9 | 38.0 | 37.9 | - |
| | 150 | 76.8 | 76.8 | 77.1 | - |
| | 210 | 137.0 | 137.2 | 137.3 | - |

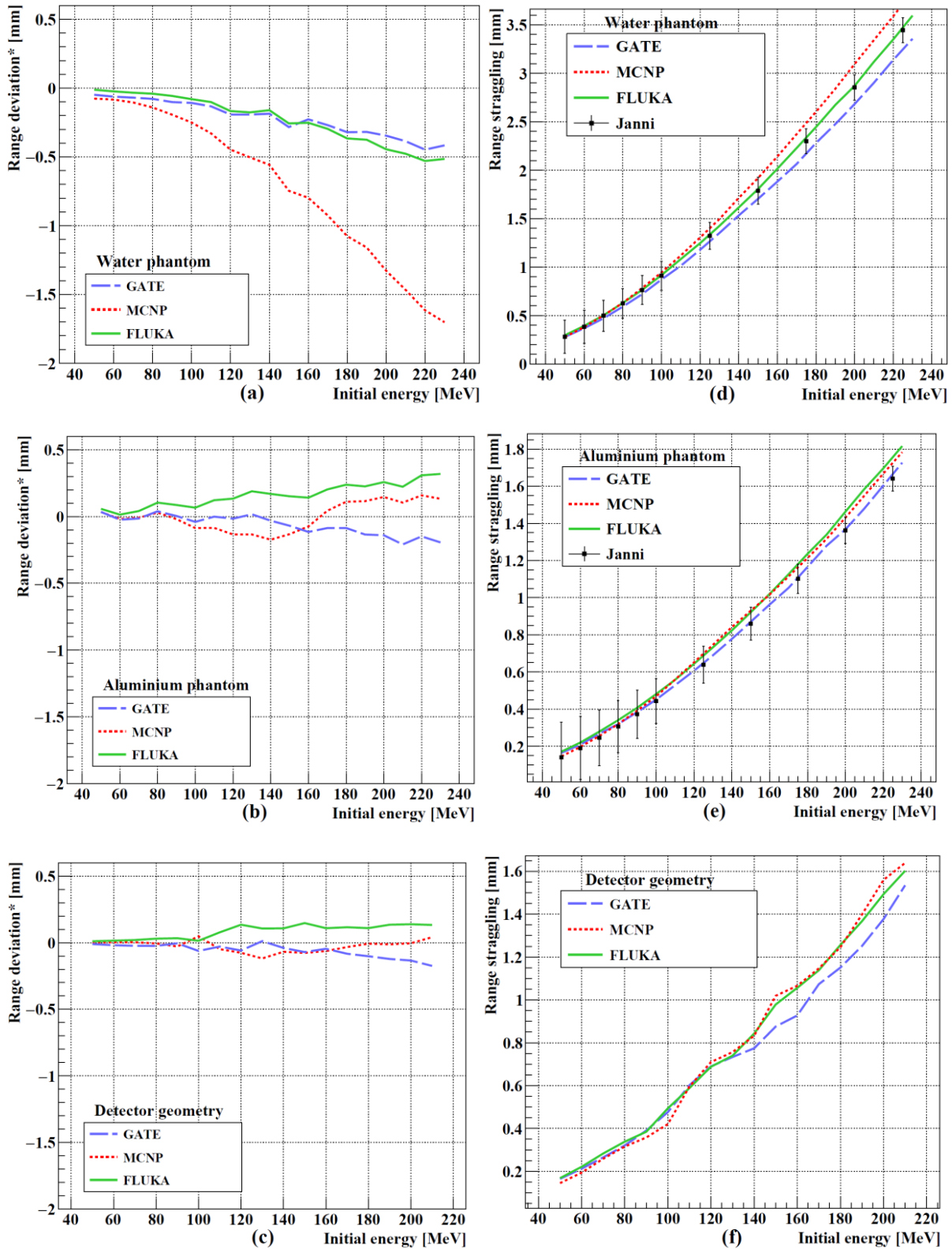

***Figure 2*** *Results for mean range deviation shown as the deviation from PSTAR (Berger et al., 2005) data as a function of the initial energy of the incoming primary protons in water (a) and aluminium (b). The range deviation for the detector geometry (c) is shown as the deviation from the average of the three MC packages. The range straggling in the same geometries is displayed in (d), (e) and (f), respectively. The corresponding semi-empirical values from J.F. Janni (Janni, 1982) are included for the water and aluminium phantoms.*

Sølie, J. R. et al. *A comparison of proton ranges in complex media using GATE/Geant4, MCNP6 and FLUKA**Table 4* MC calculated range straggling and data from J.F. Janni for 50, 100, 150 and 230 MeV primary proton energies in the water phantom, aluminium phantom and detector geometry.  *Note that the range straggling data from J.F. Janni (Janni, 1982) is for 225 MeV protons.

| Material | Energy [MeV] | GATE/Geant4 [mm] | MCNP6 [mm] | FLUKA [mm] | Janni [mm] |
|---|---|---|---|---|---|
| Water | 50 | 0.28 | 0.28 | 0.30 | 0.28 |
|  | 100 | 0.87 | 0.95 | 0.92 | 0.91 |
|  | 150 | 1.70 | 1.92 | 1.80 | 1.79 |
|  | 230 | 3.36 | 3.84 | 3.60 | 3.45* |
| Aluminium | 50 | 0.16 | 0.15 | 0.17 | 0.14 |
|  | 100 | 0.45 | 0.47 | 0.48 | 0.44 |
|  | 150 | 0.87 | 0.93 | 0.92 | 0.86 |
|  | 230 | 1.73 | 1.78 | 1.81 | 1.64* |
| Detector geometry | 50 | 0.16 | 0.14 | 0.17 | - |
|  | 100 | 0.48 | 0.42 | 0.49 | - |
|  | 150 | 0.88 | 1.02 | 0.98 | - |
|  | 210 | 1.53 | 1.64 | 1.60 | - |

*Table 5* MC calculated beam spreading for 50, 100, 150 and 230 MeV primary proton energies in the water phantom, aluminium phantom and detector geometry.

| Material | Energy [MeV] | GATE/Geant4 | MCNP6 | FLUKA |
|---|---|---|---|---|
| Water | 50 | 0.038 | 0.046 | 0.042 |
|  | 100 | 0.026 | 0.033 | 0.031 |
|  | 150 | 0.021 | 0.025 | 0.024 |
|  | 230 | 0.015 | 0.016 | 0.016 |
| Aluminium | 50 | 0.047 | 0.051 | 0.052 |
|  | 100 | 0.042 | 0.051 | 0.048 |
|  | 150 | 0.035 | 0.042 | 0.040 |
|  | 230 | 0.027 | 0.030 | 0.029 |
| Detector geometry | 50 | 0.048 | 0.051 | 0.052 |
|  | 100 | 0.044 | 0.051 | 0.048 |
|  | 150 | 0.037 | 0.042 | 0.040 |
|  | 210 | 0.031 | 0.033 | 0.032 |

## 3.3  Simulated fraction of nuclear interactions

The simulated fractions of nuclear interactions for some selected energies as well as corresponding data from J.F. Janni (Janni, 1982) are collected in **Table 6**.

The fractions of primary proton undergoing nuclear interactions are shown in **Figure 3 (d) – (f)** for water, aluminium and the detector geometry. For water and aluminium, the MC calculated results are compared to data from J.F. Janni (Janni, 1982). The fraction of nuclear interactions indicate that all three MC packages yield similar fractions of nuclear interactions, with a maximum 7.5% deviation between the MC packages in water, 6.9% in aluminium and 4.0% in the detector geometry.

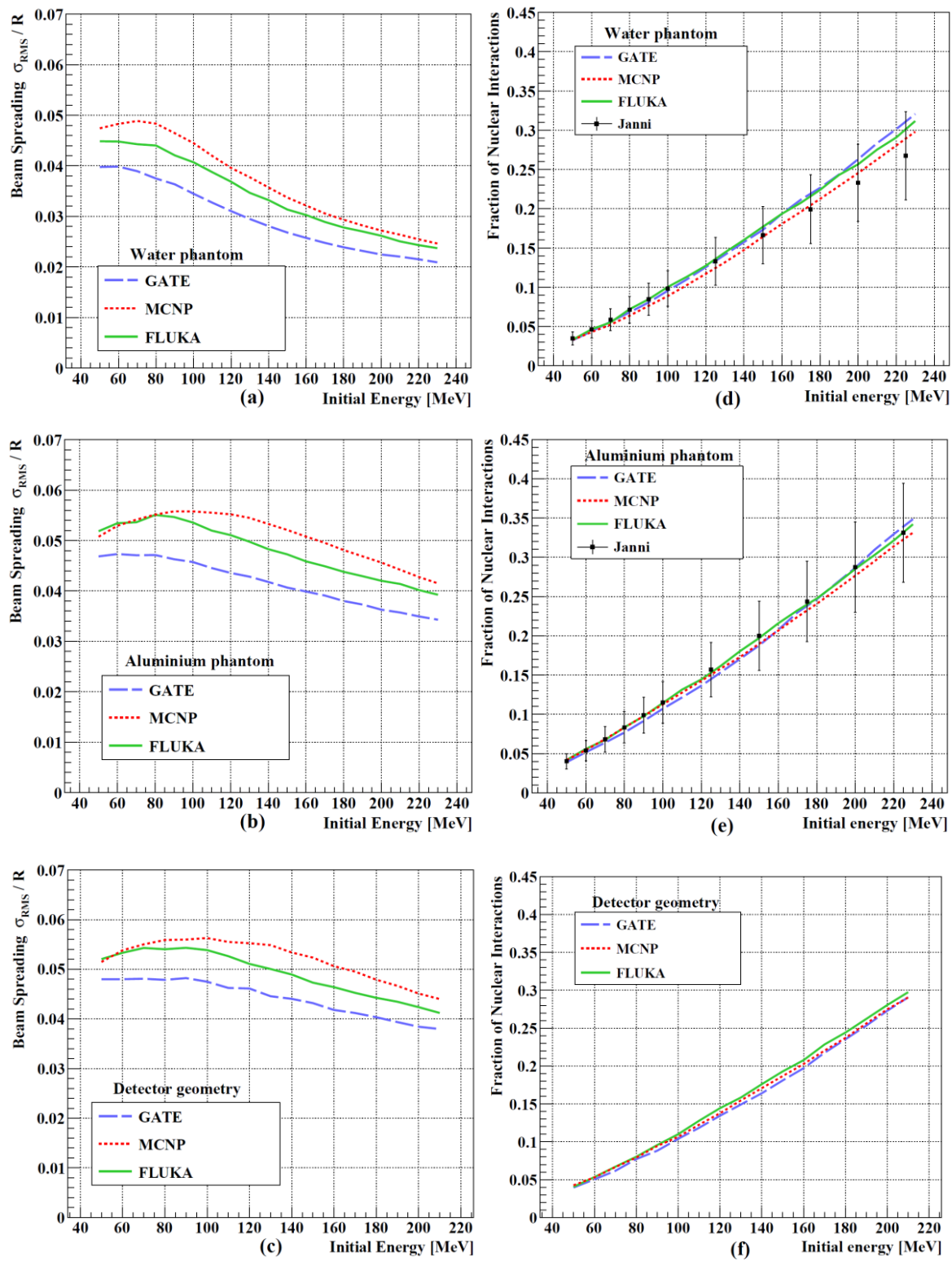

*Figure 3* Calculated beam spread in water (a), aluminium (b), and the detector geometry (c). The fraction of nuclear interactions in are displayed in (d) – (f) for the same geometries, respectively.

## 3.4  Proton beam spreading

The obtained results for beam spreading for some selected primary proton energies are listed in **Table 5**.

The complete MC simulated beam spread is shown in **Figure 3 (a) – (c)** and shows an agreement between MCNP6 and FLUKA. An overall lower amount of beam spreading can be seen in the GATE/Geant4 data compared to the other MC packages. This can also be seen in **Figure 4**, with the beam profiles of a 120 MeV proton beam incident on the water phantom from MCNP6, FLUKA and GATE/Geant4.

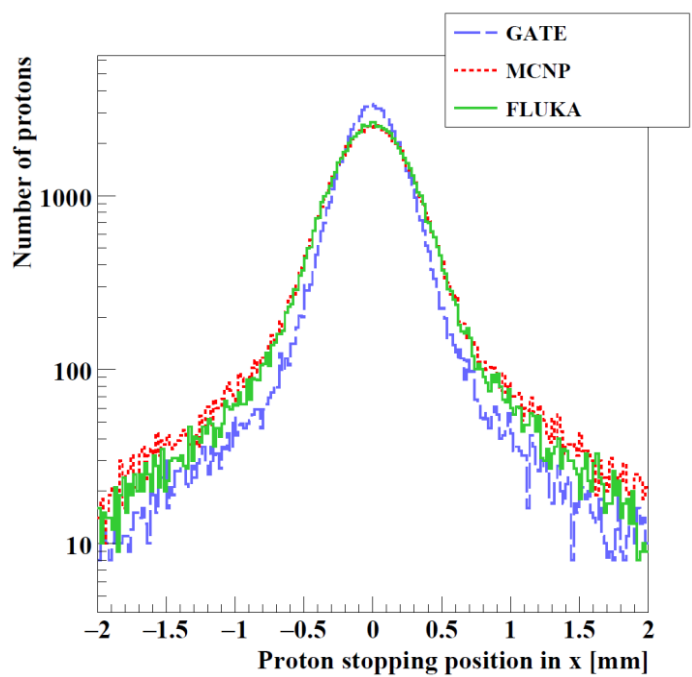

*Figure 4 the lateral profile of the Bragg Peak for the three MC software packages. The beam spreading in GATE/Geant4 is reduced compared to the other packages, similar to the trend seen in Figure 3 (a) - (c).*

*Table 6 MC calculated fraction of nuclear interactions data for 50, 100, 150 and 230 MeV primary proton energies in the water phantom, aluminium phantom and detector geometry. Also given are the corresponding data in water and aluminium published by J.F. Janni (Janni, 1982).*

| Material | Energy [MeV] | GATE/ Geant4 | MCNP6 | FLUKA | J.F. Janni |
|---|---|---|---|---|---|
| Water | 50 | 0.032 | 0.033 | 0.033 | 0.035 |
|  | 100 | 0.096 | 0.089 | 0.100 | 0.098 |
|  | 150 | 0.173 | 0.164 | 0.178 | 0.166 |
|  | 230 | 0.321 | 0.298 | 0.312 | 0.268 |
| Aluminium | 50 | 0.039 | 0.042 | 0.042 | 0.040 |
|  | 100 | 0.107 | 0.113 | 0.114 | 0.115 |
|  | 150 | 0.188 | 0.190 | 0.197 | 0.200 |
|  | 230 | 0.310 | 0.295 | 0.303 | 0.33 |
| Detector geometry | 50 | 0.040 | 0.043 | 0.041 | - |
|  | 100 | 0.104 | 0.106 | 0.110 | - |
|  | 150 | 0.181 | 0.187 | 0.193 | - |
|  | 210 | 0.291 | 0.291 | 0.297 | - |

# 4 Discussion

The objective of this study has been to compare simulated range distributions of protons traversing different materials obtained with three general purpose MC software packages, namely GATE/Geant4, MCNP6 and FLUKA, by assessing the agreement on mean projected proton range, range straggling, beam spread, and also the fraction of protons lost from the primary beam in nuclear interactions. These parameters were compared in situations with protons traversing homogeneous water and aluminium phantoms and in a proton tracking detector geometry. The MC results were also compared with data from PSTAR (Berger et al., 2005) and J.F. Janni (Janni, 1982) for particle range in water and aluminium. The main motivation for performing these comparisons was to achieve detailed knowledge about the MC packages as a validation tool in the initial design work of a proton CT detector for which experimental data is presently not available. The above-mentioned variables represent important design figures in a proton CT system concerning the range resolution and reconstruction efficiency (Pettersen et al., 2017). The mean proton ranges agree to within expected straggling, while the range straggling and fraction of nuclear interactions in water and aluminium agree with J.F. Janni published data to within uncertainties. These comparisons reveal that the largest deviation between the results presented here and the published data occur for the water phantom. Note that the range straggling values in **Figure 2 (f)** exhibits higher levels of variation dependent on the initial proton energy, compared to the homogeneous phantoms. This artefact may appear due to the fact that the range straggling, i.e. the width of the proton range distribution, depends heavily on the proton stopping position in the heterogeneous detector relative to the various sub-structures: A proton beam stopping in the air gap will have a higher straggling value compared to one stopping in the aluminium layer. For the lateral beam spreading the results are consistent with existing studies previously mentioned (Bednarz et al., 2011; Grevillot et al., 2010; Lin et al., 2017; Mertens et al., 2010).

There are however important aspects to be aware of in the planning of MC simulations and during interpretation of the results as also mentioned in a topical review article on the role of range uncertainties in MC by Paganetti (Paganetti, 2012). Awareness should be placed on how different MC packages handle the implementation of their respective models for physics interactions, which can be done either by theoretical models or through interpolation of experimental data depending on the energy region that is studied. In this regard, certain physics models and MC packages can be better suited to model a clinical proton beam than others. The recommended packages and settings in the various MC packages to be applied for this purpose, and as used in this work, are listed in **Table 1**.

User defined settings that are easily changed in one MC package, can be difficult or impossible to change in others. As is seen in **Figure 2 (a)** for the range deviation in water, MCNP6 diverges from the other packages with increasing initial proton energies. A possible cause for this divergence is the ionization potential (IP), which is an important parameter in estimating the range of protons in low Z materials (Newhauser and Zhang, 2015). Five separate GATE/Geant4 simulations with varying IP values were performed, and the resulting ranges were compared to the range predicted by MCNP6. **Figure 5** shows the range deviation between MC simulations using GATE/Geant4 and MCNP6. Note that MCNP6 uses recommended values for the IP for a material from ICRU49 (Deasy, 1994) where applicable, otherwise it uses the Bragg Additivity rule (Thwaites, 1983) to calculate the IP for composite materials. It should also be noted that by setting the IP of water to 73 eV in GATE/Geant4, instead of the ICRU49-recommended value of 75 eV (which is an often debated value that is likely to change in the future (ICRU, 2014)), the resulting proton ranges are closer to the ranges obtained in MCNP6. Other publications have also found that the results depend significantly on user defined settings in MC simulations (Kimstrand et al., 2008).

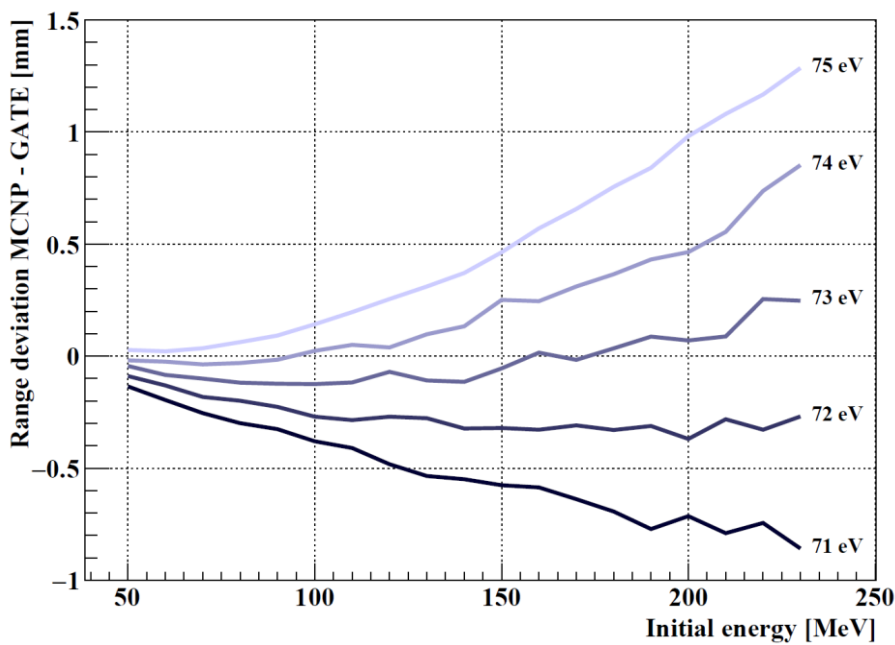

*Figure 5* Range deviation between MCNP6 and five separate simulations using GATE/Geant4 as a function of the initial proton energy. All five GATE/Geant4 simulations were performed using different values for the Ionization Potential in the 71 - 75 eV span.

# 5 Conclusion

In conclusion, although the simulated range deviation of protons is shown to increase with increasing energies, the proton range deviation between the MC software packages is sub-millimetre in the therapeutic range 50-230 MeV. The exception to this is the range predicted by MCNP6 in water which deviates by 1.2 mm and 1.3 mm from the mean proton ranges calculated by FLUKA and GATE/Geant4, respectively. A possible reason for this is the ionization potential of water used in MCNP6. Although it is reported that MCNP6 uses the recommended value of 75 eV as the ionization potential of water, five separate GATE/Geant4 simulations using different values of the ionization potential reveal that the results between GATE/Geant4 and MCNP6 agree better at an ionization potential value of about 73 eV. The same tendency with increasing proton energies is observed also in proton range straggling. However, the largest discrepancy in the predicted range straggling between the MC packages is 0.5 mm (12.5% of the range straggling value). Considering the lateral beam spread, MCNP6 reports an overall higher amount of beam spread, and GATE/Geant4 an overall lower amount than the other packages, however, they are observed to converge at higher proton energies. Similar discrepancies in beam spreading is reported in previous publications on the subject matter (Kimstrand et al., 2008; Lin et al., 2017; Mertens et al., 2010). The MC calculated fraction of nuclear interactions compares relatively well with data from J.F. Janni (Janni, 1982) and the MC packages show a maximum deviation of 7.5% from each other in the case of water at proton energy of 230 MeV. Due to the general agreement between the output values from the different MC software packages, the choice of simulation framework may be made on personal preferences or inter-project compatibility.

# Acknowledgements

This work was supported by the Bergen Research Foundation (93116) and by the Western Norway Regional Health Authority (911933).